\documentclass[11pt]{article}
\pdfoutput=1
\usepackage{jheppub}
\usepackage{amsmath}
\usepackage{comment}

\newlength{\dummysp}
\newcommand{\tr}{\mathop{{\hbox{Tr} \, }}\nolimits}

\newcommand{\beq}{\begin{eqnarray}}
\newcommand{\eeq}{\end{eqnarray}}
\newcommand{\nnn}{ \nonumber \\ }
\newcommand{\ddd}{ \nnn && }
\newcommand{\p}{{\partial}}
\newcommand{\e}{{\epsilon}}

\newcommand{\ord}[1]{{{\cal O}(#1)}}
\newcommand{\gappeq}{\mathrel{\rlap {\raise.5ex\hbox{$>$}}
{\lower.5ex\hbox{$\sim$}}}}
\newcommand{\lappeq}{\mathrel{\rlap{\raise.5ex\hbox{$<$}}
{\lower.5ex\hbox{$\sim$}}}}
\newcommand{\myref}[1]{(\ref{#1})}

\newcommand{\ben}{\begin{enumerate}}
\newcommand{\een}{\end{enumerate}}
\newcommand{\bit}{\begin{itemize}}
\newcommand{\eit}{\end{itemize}}
\newcommand{\Cbf}{{\bf C}}

\newcommand{\Ncal}{{\cal N}}

\newcommand{\psib}{{\bar \psi}}

\newcommand{\Zbf}{{\bf Z}}

\newcommand{\cA}{{\cal A}}
\newcommand{\cAb}{{\overline{\cal A}}}
\newcommand{\cF}{{\cal F}}
\newcommand{\cFb}{{\overline{\cal F}}}
\newcommand{\cD}{{\cal D}}
\newcommand{\cDb}{{\overline{\cal D}}}
\newcommand{\cQ}{{\cal Q}}
\newcommand{\cU}{{\cal U}}

\newcommand{\cUb}{{\overline{\cal U}}}

\newcommand{\etab}{{\overline{\eta}}}
\newcommand{\phib}{{\overline{\phi}}}

\def\[{\left [}
\def\]{\right ]}
\def\({\left (}
\def\){\right )}

\title{Real space renormalization group for twisted lattice ${\cal N}=4$ super Yang-Mills}

\author[a]{Simon Catterall,}
\author[b]{Joel Giedt} 

\affiliation[a]{Department of Physics, Syracuse University, 
Syracuse, New York, 13244 USA}
\affiliation[b]{Department of Physics, Applied Physics, and Astronomy, \\ 
Rensselaer Polytechnic Institute, 110 8th St., Troy, New York, 12180, USA}

\emailAdd{smcatterall@gmail.com}
\emailAdd{giedtj@rpi.edu}

\abstract{
A necessary ingredient for our previous results on the form of the long distance effective
action of the twisted lattice ${\cal N}=4$ super Yang-Mills theory is the existence of a
real space renormalization group which preserves the lattice structure, both the symmetries
and the geometric interpretation of the fields.  In this brief article we provide an
explicit example of such a blocking scheme and illustrate its practicality in the context of a small scale
Monte Carlo renormalization group calculation.  We also discuss the implications of this
result, and the possible ways in which to use it in order to obtain further information
about the long distance theory. 
}
  
\begin{document}
\maketitle
\flushbottom

\section{Introduction}
There has been significant recent progress on the lattice discretization of $\Ncal=4$ 
super Yang-Mills (SYM) \cite{Kaplan:2005ta,Catterall:2007kn,Catterall:2011pd,Catterall:2013roa,
Catterall:2014vka}.  (For alternative approaches see \cite{Ishii:2008ib, Ishiki:2008te,Ishiki:2009sg, Hanada:2010kt, Honda:2011qk,Honda:2013nfa}.)  One motivation for such
efforts is that it is highly desirable to test the AdS/CFT correspondence at a finite
number of colors $N$, and for moderate values of the 't Hooft coupling $\lambda = g^2 N \sim 1$.
Indeed, results in this regime would, in theory, open the way to nonperturbative results
for quantum gravity.  Another reason to study $\Ncal=4$ SYM on the lattice is that
the continuum theory is an interacting conformal field theory at all scales, unlike
the situation with  theories inside the 
conformal window, which only approach a conformal fixed point in the infrared (IR).

The key new idea which underlies these new lattice constructions is to discretize not the usual
theory but a {\it topologically twisted}
cousin. In flat space this corresponds merely to an exotic change of variables --- one more suited to
discretization.  In the case of $\Ncal=4$ SYM there are three independent topological
twists of the theory and the one that is employed in the lattice work is the Marcus or
Geometric-Langlands twist \cite{Marcus:1995mq,Kapustin:2006pk}. The resulting lattice action takes the form

\beq
S &=&  \frac{1}{2g^2} ( \cQ \lambda + S_{\text{closed}} ) \nnn
\lambda &=& \sum_x a^4 \tr ( \chi_{ab} {\cal F}_{ab} + \eta \overline{\cal D}_a^{(-)} \cU_a
- \frac{1}{2} \eta d ) \nnn
S_{\text{closed}} &=& -\frac{1}{4} \sum_x a^4 \e_{abcde} \chi_{de} \overline{\cal D}_c^{(-)} \chi_{ab}(x)
\eeq
where we include the appropriate factors of the lattice spacing $a$ 
and the explicit expressions for the terms involving covariant derivatives are given by
\beq
{\cal F}_{ab}(x) &=& {\cal D}_a^{(+)} {\cal U}_b(x) 
= {\cal U}_a(x) {\cal U}_b(x+e_a) - {\cal U}_b(x) {\cal U}_a(x+e_b) \nnn
\overline{\cal D}_a^{(-)} {\cal U}_a(x)
&=& {\cal U}_a(x) \overline{\cal U}_a(x) - \overline{\cal U}_a(x-e_a) {\cal U}_a(x-e_a) \nnn
\e_{abcde} \chi_{de} \overline{\cal D}_c^{(-)} \chi_{ab}(x)
&=& \e_{abcde} \chi_{de}(x+e_a+e_b) [ \chi_{ab}(x) \overline{\cal U}_c(x-e_c)
\ddd \quad - \overline{\cal U}_c(x-e_c+e_a+e_b) \chi_{ab}(x - e_c) ]
\eeq
Notice that these expressions involve fields which are associated to the links of an $A_4^*$ lattice which possesses
five (linearly dependent) basis vectors  and an associated $S^5$ point group symmetry.
To complete the specification of the action we also need the action of $\cQ$ on
the lattice fields, which is given by
\beq
&& \cQ {\cal U}_a = \psi_a, \quad \cQ \psi_a = 0, \quad \cQ \overline{\cal U}_a = 0 \nnn
&& \cQ \chi_{ab}(x) = -\overline{\cal F}_{ab}(x) 
\equiv \overline{\cal U}_b(x+e_a) \overline{\cal U}_a(x) 
- \overline{\cal U}_a(x+e_b) \overline{\cal U}_b(x) \nnn
&& \cQ \eta = d, \quad \cQ d =0
\eeq
It can be checked that the classical continuum limit of this lattice action yields the usual
Marcus twist of $\Ncal=4$ SYM if the lattice fields are decomposed into their irreducible components under
the $S^5$ symmetry (see \cite{Unsal:2006qp})  and the link fields  expanded according to\footnote{We
work with antihermitian generators of the SU(N) gauge group.}
\beq
\cU_a(x) = \frac{1}{a} + \cA_a(x), \quad \cUb_a(x) = \frac{1}{a} - \cAb_a(x)
\label{linkexpansion}
\eeq
As an example of this argument consider the $A_4^*$ term $\sum \chi_{ab}\cD_{\left[a\right.}\psi_{\left.b\right]}$ which emerges after carrying out
the $\cQ$-variation of the $\chi_{ab}\cF_{ab}$ term above.  Decompose the lattice fields (in a fixed gauge) into their $S^5$ irreducible components
via the relations
\begin{eqnarray}
\chi_{ab}&=&P_{a\mu}P_{b\nu}\chi_{\mu\nu}+P_{a5}P_{b\nu}\psib_\nu\\
\psi_b&=&P_{b\lambda}\psi_\lambda+P_{b5}\etab\\
\cD_a&=&P_{a\rho}\cD_\rho+P_{a5}\phi
\end{eqnarray}
The $5\times 5$ orthogonal matrix $P$ that appears in these expressions is introduced in \cite{Unsal:2006qp} and serves as the bridge between the fields occurring on the $A_4^*$ lattice
and their continuum cousins.\footnote{It is crucial for these arguments that in fact the lowest lying irreducible representations of the $S^5$ (strictly its  $A^5$ subgroup)
match those of the continuum twisted $SO(4)$ group} In these expressions Greek indices run from one to four while Latin cover the range from
one through five. If we substitute these decompositions into this $A_4^*$ fermion term we find
\begin{eqnarray}
\chi_{ab}\cD_a\psi_b&=&P_{a\mu}P_{b\nu}P_{a\rho}P_{b\lambda}\chi_{\mu\nu}\cD_\rho\psi_\lambda+P_{a5}P_{b\nu}P_{b\lambda}P_{a5}\psib_\nu\phi\psi_\lambda+\ldots\\
\chi_{ab}\cD_b\psi_a&=&P_{a\mu}P_{b\nu}P_{b\rho}P_{a\lambda}\chi_{\mu\nu}\cD_\rho\psi_\lambda+P_{a5}P_{b\nu}P_{a5}P_{b\rho}\psib_\nu \cD_\rho\etab+\ldots
\end{eqnarray}
Using the orthogonal properties of the matrix $P$ all other terms vanish since they involve contractions of the type $P_{a\mu}P_{a5}=0$ and the expression
simplifies to
\beq
\sum \chi_{ab}\cD_{\left[a\right.}\psi_{\left.b\right]}\to \chi_{\mu\nu}\cD_{\left[\mu\right.}\psi_{\left.\nu\right]}+\psib_\mu\cD_\mu \etab+\psib_\nu[\phi,\psi_\nu]
\label{DK}\eeq
These terms  match precisely some of those appearing in the continuum Marcus twist of ${\cal N}=4$ SYM.  Similar reductions occur for all terms in the $A_4^*$ action
and confirm that the lattice theory does indeed target ${\cal N}=4$ SYM in the naive continuum limit.
Notice that the action has an additional $U(1)$ ghost number\footnote{This is
referred to as a ghost number because $\cQ$ is used as a BRST
symmetry in the construction of a topological
field theory from the Marcus twist.} symmetry under which  the fields $(\eta,\psi,\chi,\psib,\etab,\phi,\phib,\cA)$ carry charges $(1,-1,1,-1,1,2,-2,0)$. This symmetry is hidden in the original $A_4^*$ lattice formulation and is visible only when the $A_4^*$ fields are decomposed into their
irreducible representations under the $S^5$ lattice symmetry. It will be important in our later analysis.

\section{Blocking transformation}
\label{sect:blocking}
The original lattice $\Lambda$ may be described by 
$\Lambda = \{ a \sum_{\mu=1}^4 n_\mu e_\mu | n \in \Zbf^4 \}$,
where the $e_\mu$ are the first four of the five (degenerate) basis vectors of the $A_4^*$ lattice.
The blocked lattice will merely be doubled in every direction:
$\Lambda' = \{ 2 a \sum_{\mu=1}^4 n_\mu e_\mu | n \in \Zbf^4 \}$.  From this point
forward we will work in lattice units, setting $a=1$.
The blocked fields will be denoted by primes and must begin and end on sites of the
blocked lattice $\Lambda'$.  The trick is to come up with a blocking transformation
such that the $\cQ$ algebra is preserved (maintenance of $S_5$ symmetry will be
straightforward), with the geometric intepretation also surviving.  For example, $\chi'_{ab}(x)$
must begin on site $x + 2e_a + 2e_b$ and end on site $x$ since the
original field $\chi_{ab}(x)$ begins on $x+e_a+e_b$ and ends on $x$.  One choice that achieves this
is the following:
\beq
\cU'_a(x) &=& \xi \cU_a(x) \cU_a(x+e_a), \quad \cUb'_a(x) = \xi \cUb_a(x+e_a) \cUb_a(x) \nnn
d'(x) &=& \xi d(x), \quad \eta'(x) = \xi \eta(x) \nnn
\psi'_a(x) &=& \xi [ \psi_a(x) \cU_a(x+ e_a) + \cU_a(x) \psi_a(x + e_a) ] \nnn
\chi'_{ab}(x) &=& \frac{\xi}{2} [ \cUb_a(x+e_a+2e_b) \cUb_b(x+e_a+e_b) \chi_{ab}(x) 
+ \cUb_b(x+2e_a+e_b) \cUb_a(x+e_a+e_b) \chi_{ab}(x) ]
\ddd + \xi [ \cUb_a(x+e_a+2e_b) \chi_{ab}(x+e_b) \cUb_b(x) 
+ \cUb_b(x+2e_a+e_b) \chi_{ab}(x+e_a) \cUb_a(x) ]
\ddd + \frac{\xi}{2} [ \chi_{ab}(x+e_a+e_b) \cUb_a(x+e_b) \cUb_b(x) 
+ \chi_{ab}(x+e_a+e_b) \cUb_b(x+e_a) \cUb_a(x) ]
\label{block}
\eeq
Because the link variables, being elements of $GL(N,\Cbf)$, are non-compact
we have allowed for the possibility that they are rescaled by a
factor $\xi$ under the transformation.  (This will become important
when we perform the two-lattice matching in our Monte Carlo renormalization
group (MCRG) analysis of Section \ref{mcrg}, and in this context $\xi$ becomes
a blocking parameter similar to those in other schemes.  Indeed, it is typical
in MCRG to tune a blocking parameter in order to achieve matching.)
The following parts of the algebra are obvious upon inspection:  $\cQ \cUb'=0$, $\cQ \eta' = d'$,
$\cQ d' = 0$.  In particular note that for the $\eta,d$ system we have simply utilized
decimation.  It is not difficult to also see that $\cQ \cU'_a = \psi'_a$ by making
use of the orginal algebra $\cQ \cU_a = \psi_a$.  The fact that $\cQ \psi'_a = 0$
then follows from the minus sign that comes in when $\cQ$ is pushed past $\psi_a(x)$:
\beq
\cQ \psi'_a(x) &=& \xi \cQ [ \psi_a(x) \cU_a(x+ e_a) + \cU_a(x) \psi_a(x + e_a) ]
\ddd = -\xi \psi_a(x) \psi_a(x+e_a) + \xi \psi_a(x) \psi_a(x+e_a) = 0
\eeq
To demonstrate that $\cQ \chi'_{ab} = -\cFb'_{ab}$ we first note that the logical
definition of the field strength in terms of the blocked fields is a straightforward
transcription of the original expression:
\beq
\cFb'_{ab}(x) = -\xi [ \cUb'_b(x+2e_a) \cUb'_a(x) - \cUb'_a(x+2e_b) \cUb'_b(x) ]
\eeq
Then applying $\cQ$ to the expression for $\chi'_{ab}$ in terms of the original
fields, one indeed obtains the desired expression after a few steps of algebra.
At this point one immediately recognizes that the nilpotency $\cQ^2=0$ has
also been maintained.  It is also easy to see that the properties under the
symmetric group $S^5$ have been preserved:  any invariant of the original
fields is also $S^5$ invariant when expressed in terms of the blocked fields.
For instance, $\sum_a \cU'_a \cUb'_a$ is obviously invariant under
permutations of the indices.

\section{Renormalization}
\label{sect:baction}
What we are interested in is the number of counterterms that must be fine-tuned
in order to obtain the desired long distance effective theory---i.e., one
whose classical continuum limit is nothing but ${\cal N}=4$ SYM.
The strategy is to enumerate the lattice operators that could possibly
be generated under renormalization group flow with the blocking
scheme given above.  Lattice operators that give relevant or marginal
operators in the continuum limit are the ones that would correspond
to counterterms which must be fine-tuned.  Of course some operators
can be given their canonical coefficients simply by a rescaling of
the fields; this is something that we will also describe below.  The
remaining coefficients, which are determined by the flow from the
ultraviolet theory (UV), would have to be fine-tuned by adjusting corresponding
coefficients in that UV theory.  If we can write down two lattice operators
that both give the same relevant/marginal operator and only differ
by irrelevant operators in the continuum limit, then we can count
them as a single counterterm for the purpose of fine-tuning, and
we only need write one of them for our description of the ``most general
long distance effective action.''  This is because this long distance
action is defined up to irrelevant operators, which do not affect
the counting of counterterms that must be fine-tuned.

In the continuum theory, the $\cQ$ closed term that appears in the action is the unique
renormalizable operator with this property.  Hence we know that on the lattice the
$\cQ$ closed term is also unique.  Thus what remains is to enumerate the $\cQ$
exact operators that are renormalizable.  These must all take the form $\cQ \tr [ \Psi f(\cU,\cUb,d) ]$
or $\cQ \{ \tr \eta \tr f(\cU, \cUb,d) \}$,
where $\Psi$ is one of the fermion fields.  Cubic or higher powers of fermions
would be nonrenormalizable, and the quantity that $\cQ$ acts on must be fermionic
so that the action is bosonic.  Only $\eta$ can be used in a double trace
operator, because a field must be a site field in order for its trace
to be gauge invariant.  Thus, beginning with $\Psi = \eta$, we have the following
possible terms:
\beq
&& \cQ \tr [ \eta(x) \overline{\cal U}_a(x-e_a) {\cal U}_a(x-e_a) ], \quad \cQ \tr ( \eta d )
\ddd \cQ \tr [ \eta(x) {\cal U}_a(x) \overline{\cal U}_a(x) ], \quad Q {\rm Tr} \eta, 
\ddd \cQ [ \tr \eta \tr ( {\cal U}_a \overline{\cal U}_a ) ]
\eeq
However, the original action is invariant under the shift symmetry
\beq
\eta(x) \to \eta(x) + c {\bf 1}_N
\eeq
where $c$ is an arbitrary constant Grassmann parameter.  This symmetry restricts
the above terms to the following combinations:
\beq
&& \cQ \tr [ \eta \overline{\cal D}_a^{(-)} {\cal U}_a ], \quad \cQ \tr ( \eta d )
\ddd \cQ \tr ( \eta {\cal U}_a \overline{\cal U}_a ) 
- \frac{1}{N} \cQ \{ \tr \eta \tr ( {\cal U}_a \overline{\cal U}_a ) \}
\eeq
Thus we only find one term that is not already present in the original action;
this will be the so-called ``$\beta$ term'' below.  

As far as $\Psi = \psi_a$ is
concerned, one gauge invariant combination that we can write down is
$\tr \psi_a \cUb_a$.  However, $\cQ$ acting on this vanishes identically.
Another operator that is allowed by the symmetries is
\beq
\Delta S = \beta_2 \sum_x a^4 ( a \cQ \tr \psi_a \cU_a \cUb_a \cUb_a )
= \beta_2 \sum_x a^4 [ a \tr \psi_a(x) \psi_a(x+e_a) \cUb_a(x+e_a) \cUb_a(x) ]
\eeq
where $\beta_2$ is a dimensionless constant generated under the renormalization group (RG) flow,
and the power of $a$ in front of the operator is dictated by the mass
dimensions of the fields, according to the way in which we normalized
the links in \myref{linkexpansion}:  $[ \psi_a ] = 3/2$, $[\cU_a]=1$.
Of course the factor of $a^4$ simply represents the measure $d^4 x$ in
the continuum limit, as in the original action above.  The explicit power
of $a$ in front of the operator makes it appear as if this is an
irrelevant operator in the continuum limit; however, this is not the
case because of the factors of $1/a$ that arise from \myref{linkexpansion}.
Explicitly:
\beq
&& a \tr \psi_a(x) \psi_a(x+e_a) \cUb_a(x+e_a) \cUb_a(x)
\ddd \qquad = a \tr \bigg\{ \psi_a(x) [ \psi_a(x) + a \p_a \psi_a(x) + \ord{a^2} ]
\[ \frac{1}{a} - \cAb_a(x) + \ord{a} \] \[ \frac{1}{a} - \cAb_a(x) \] \bigg\}
\ddd \qquad = \tr \psi_a \overline{\cal D}_a \psi_a + \ord{a}
\eeq
Thus at leading order there is a marginal operator coming from this
term.  It violates the Euclidean $SO(4)$ Lorentz symmetry, but is
consistent with the $S^5$ point group symmetry of the lattice. However, in fact this
operator is prohibited by the  $U(1)$ symmetry described earlier and hence $\beta_2=0$
in the renormalized theory.

For the fermion choice of $\Psi=\chi_{ab}$, we can form the operators
\beq
\cQ \tr ( \chi_{ab} \cU_a \cU_b ), \quad \cQ \tr ( \chi_{ab} \cU_b \cU_a )
\eeq
However, the antisymmetry $\chi_{ab} = -\chi_{ba}$ requires that these be
combined with a minus sign, leading to the operator
\beq
\cQ \tr ( \chi_{ab} {\cal D}_a^{(+)} {\cal U}_b )
\label{chieq}
\eeq
which is already present in the action.  As before, adding additional powers
of $\cU_a \cUb_a$ merely leads to the same marginal operator in the continuum
limit; leaving them out only changes irrelevant operators---something that
we are not interested in as far as counting counterterms is concerned.

It is clear that the blocked fields must have the same
geometric interpretation on the lattice $\Lambda'$ in order
for these arguments to hold.  This dictates the structure of the
site arguments of the fields, for instance appearing in \myref{chieq},
such that the same term as in the original action appears in
the long distance effective theory.  It is also important
that the blocking preserves the $S^5$ symmetry, so that
this restriction on operators will be present.  Without it,
we would have generated many other possibilities in the above
analysis.

Thus the most general long distance effective action is\footnote{Actually there is one further operator that can be added to
the $A_4^*$ action ${\cal O}=\sum_x\epsilon_{abcde}{\rm Tr}\left(\cUb_a(x)\cUb_b(x+a)\cUb_c(x+a+b)\cUb_d(x+a+b+c)\cUb_e(x+a+b+c+d)\right)$. However this
operator (which is $\cQ$ exact) yields only the usual topological term $\int \epsilon_{\mu\nu\rho\lambda} F_{\mu\nu}F_{\rho\lambda}$ in the continuum limit.}
\beq
&& \cQ \tr \{ \alpha_1 \chi_{ab} \cF_{ab} + \alpha_2 \eta [ \cDb_a , \cD_a ]
- \frac{\alpha_3}{2} \eta d \} - \frac{\alpha_4}{4} \epsilon_{abcde} \tr \chi_{de}
\cDb_c \chi_{ab} 
\ddd \qquad + \beta Q \{ \tr \eta \cU_a \cUb_a - \frac{1}{N} \tr \eta \tr \cU_a \cUb_a \}
\eeq
where we have suppressed an overall $\sum_x a^4$ factor. 
Acting with $\cQ$, followed by a rescaling of fields
\beq
\eta \to \lambda_\eta \eta, \quad \chi_{ab} \to \lambda_\chi \chi_{ab}, \quad
\psi_a \to \lambda_\psi \psi_a, \quad d \to \lambda_d d
\eeq
we obtain
\beq
&& \tr \big\{ -\alpha_1 \cFb_{ab} \cF_{ab} - \alpha_1 \lambda_\chi \lambda_\psi \chi_{ab} \cD_{[a} \psi_{b]}
+ \alpha_2 \lambda_d d [ \cDb_a , \cD_a ] - \alpha_2 \lambda_\eta \lambda_\psi \eta \cDb_a \psi_a
\ddd -\frac{\alpha_3}{2} \lambda_d^2 d^2 - \frac{\alpha_4}{4} \lambda_\chi^2 \epsilon_{abcde}
\chi_{de} \cDb_c \chi_{ab} \big\} + \beta \big\{ \lambda_d \tr ( d \cU_a \cUb_a ) - \lambda_\eta
\lambda_\psi \tr ( \eta \psi_a \cUb_a )
\ddd -\frac{1}{N} \lambda_d \tr d \tr ( \cU_a \cUb_a ) + \frac{1}{N} \lambda_\eta \lambda_\psi
\tr \eta \tr ( \psi_a \cUb_a ) \big\}
\eeq
Using the freedom in the four rescaling factors, we can simultaneously impose four constraints,
\beq
\alpha_1 \lambda_\chi \lambda_\psi = \alpha_1, \quad \alpha_2 \lambda_d = \alpha_1,
\quad \alpha_2 \lambda_\eta \lambda_\psi = \alpha_1, \quad \alpha_4 \lambda_\chi^2 = \alpha_1
\eeq
which sets many of the coefficients above to $\alpha_1$.  Solving this system one obtains
\beq
\lambda_\eta = \sqrt{\frac{\alpha_1^3}{\alpha_4 \alpha_2^2}}, \quad
\lambda_\chi = \frac{1}{\lambda_\psi} = \sqrt{\frac{\alpha_1}{\alpha_4}}, \quad
\lambda_d = \frac{\alpha_1}{\alpha_2}
\eeq
It is also convenient to define
\beq
\alpha'_3 = \alpha_3 \( \frac{\alpha_1}{\alpha_2} \)^2, \quad
\beta' = \beta \frac{\alpha_1}{\alpha_2}
\eeq
Then the action takes the form
\beq
&& \tr \big\{ -\alpha_1 \cFb_{ab} \cF_{ab} - \alpha_1 \chi_{ab} \cD_{[a} \psi_{b]}
+ \alpha_1 d [ \cDb_a , \cD_a ] - \alpha_1 \eta \cDb_a \psi_a
\ddd -\frac{\alpha'_3}{2} d^2 - \frac{\alpha_1}{4}  \epsilon_{abcde}
\chi_{de} \cDb_c \chi_{ab} \big\} + \beta' \big\{ \tr ( d \cU_a \cUb_a ) - \tr ( \eta \psi_a \cUb_a )
\ddd -\frac{1}{N} \tr d \tr ( \cU_a \cUb_a ) + \frac{1}{N} 
\tr \eta \tr ( \psi_a \cUb_a ) \big\}
\eeq
In fact it is remarkable that the $\beta$ term does not bifurcate into multiple coefficients; this
is a consequence of $\lambda_d = \lambda_\eta \lambda_\psi$.

At this point, after rescaling of the fields, we find that a total of at most two fine-tunings will
be required:  $\alpha'_3 \to \alpha_1$ and $\beta' \to 0$.  (The overall factor of
$\alpha_1$ just corresponds to the renormalized gauge coupling, which does
not need to be fine-tuned since the continuum theory is conformal.)  This is drastically superior
to the case of a naive implementation such as Wilson fermions with SO(4) symmetry (the SO(6)
symmetry cannot be preserved because it is chiral), where there are eight fine-tunings
(see Appendix \ref{wilson}.).

However, there is another tool at our disposal.  In \cite{Catterall:2011pd} it was
shown that no effective potential was generated for the bosonic fields at any order
in lattice perturbation theory.  Thus the moduli space is not lifted by perturbative
radiative corrections.  If this is also true of nonperturbative effects, then
the $\beta$ term is forbidden, since it includes trilinear coupling of the scalars,
which would lift the moduli space (see Appendix \ref{beta}).  This would mean that under the RG flow,
$\beta \equiv 0$ is maintained.  Any deviation from this would have to arise
from nonperturbative phenomena.  It would be interesting to study the effects of
instantons in the lattice theory in order to see whether or not they generate
an effective potential.

These arguments reveal that a {\it single fine tuning of a marginal operator} $c2=\alpha'_3/\alpha_1$ is all that should be required to target 
${\cal N}=4$ SYM in the continuum limit defined by $L\to\infty$ with $g^2$ held fixed.
The situation is similar to the case of lattice QCD with Wilson fermions
where the bare mass must be fine-tuned to achieve the chiral limit.  In actuality
our situation is somewhat better because we do not need to tune the bare
coupling in order to achieve the desired lattice spacing.  This is a
consequence of the fact that the continuum theory is conformal at all scales.

We should also comment on our recent work \cite{Catterall:2013roa} 
involving the restoration of R symmetries,
which in the continuum is a global $SU(4)$ symmetry that does not commute with
supersymmetry.  It was found that restoration of even a discrete
version of the R symmetry, denoted by $R_a$ and $R_{ab}$, is sufficient
to guarantee the correct continuum limit.  It has the effect of setting
$\beta \equiv 0$ and all of the $\alpha_i$ coefficients equal to
each other.  Thus in a Monte Carlo renormalization group  analysis (see next section)
using the above blocking scheme, it should be seen that blocked observables
are related to each other by $R_a$ symmetry after a sufficient number
of steps.  The $R_a$ symmetry was tested for $1 \times 1$ Wilson loops
in \cite{Catterall:2014vka} and it was found to be violated by $\ord{10}$\%.
It would be of interest to repeat this measurement after a few
blocking steps and check whether or not the violation is reduced.

\section{Monte Carlo renormalization group}
\label{mcrg}
The strategy here is in principle relatively simple, though in practice rather challenging.
One simulates the theory on a fine lattice of size $L^4$, obtaining configurations of the fine
lattice fields.  This ensemble of fine lattice fields is then blocked according to the
procedure outlined in Section \ref{sect:blocking} to produce an ensemble of blocked lattices
of size $(L/2)^4$. 
Observables are then computed from these
blocked fields.  These could
include $ m \times  n$ Wilson loops, or mesonic correlation functions using
blocked fermions in the interpolating operators.  The so-called ``gluino-glue'' state
would also be of interest.  One then simulates a coarse lattice of size $(L/2)^4$, but with the
more general action given in the preceding section with a single additional coupling $c2$ (we set $\beta=\beta_2=0$ following
the arguments in Section~\ref{sect:baction}).
The same class of observables are now computed directly on the coarser lattice.  For instance,
if an $m \times n$ Wilson loop was computed on the blocked lattice, then a $m \times n$
Wilson loop is computed on the coarse lattice.  The coupling $c2$
of the coarse lattice action is then  tuned until there is a match
between the observables.  Notice that this matching is done at the same lattice volume and hence the leading finite
size effects are removed.
This gives one MCRG step.  Similar to more conventional MCRG blocking schemes used for
lattice QCD we have an adjustable blocking parameter, the scaling factor $\xi$,
that can be tuned to optimize the matching
between different observables.

\begin{figure}
\begin{center}
\includegraphics[width=4in,height=6in,angle=-90]{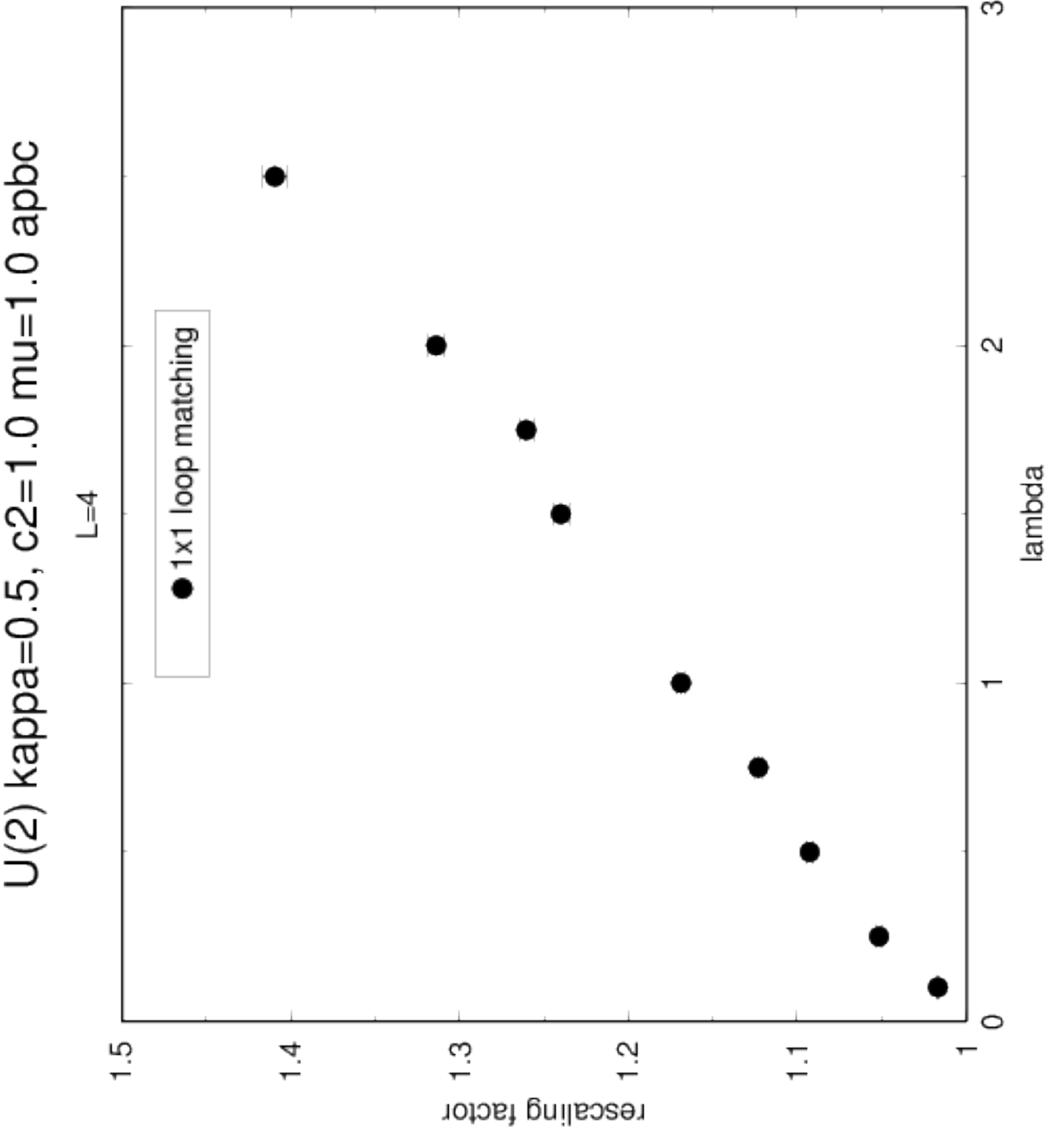}
\caption{Determination of the scaling parameter $\xi$.  Plotted on the
vertical axis is $\xi^4$, the rescaling factor needed to match the $1\times 1$ Wilson loop measured on the blocked lattices to its
value on the coarse lattice.  \label{rescale} }
\end{center}
\end{figure}

\begin{figure}
\begin{center}
\includegraphics[width=4in,height=6in,angle=-90]{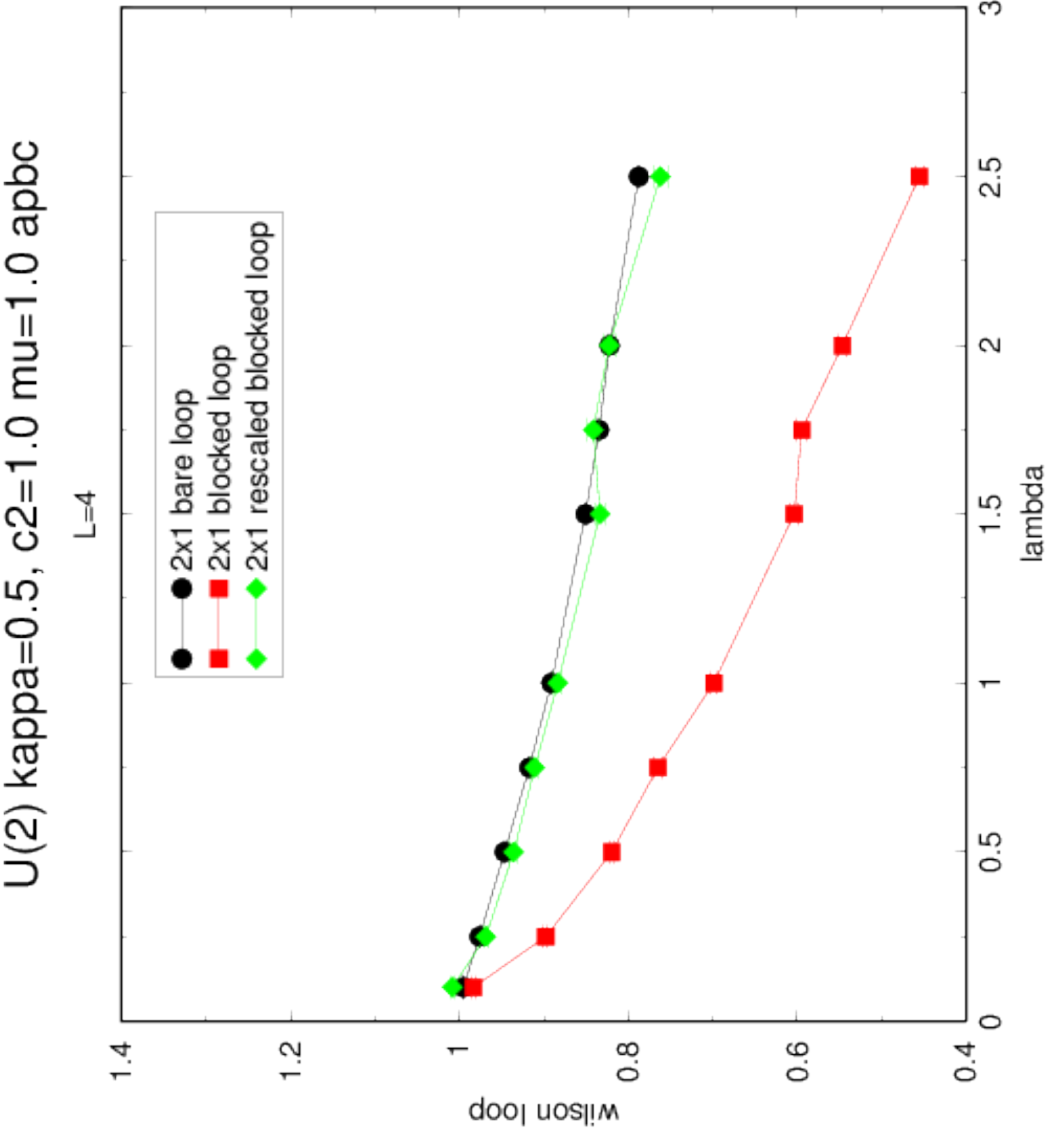}
\caption{A comparison of $W(2,1)_{b.f.}$ and $W(2,1)_{\text{coarse}}$ with
the rescaling factor taken into account.  \label{2b1} }
\end{center}
\end{figure}

\begin{figure}
\begin{center}
\includegraphics[width=4in,height=6in,angle=-90]{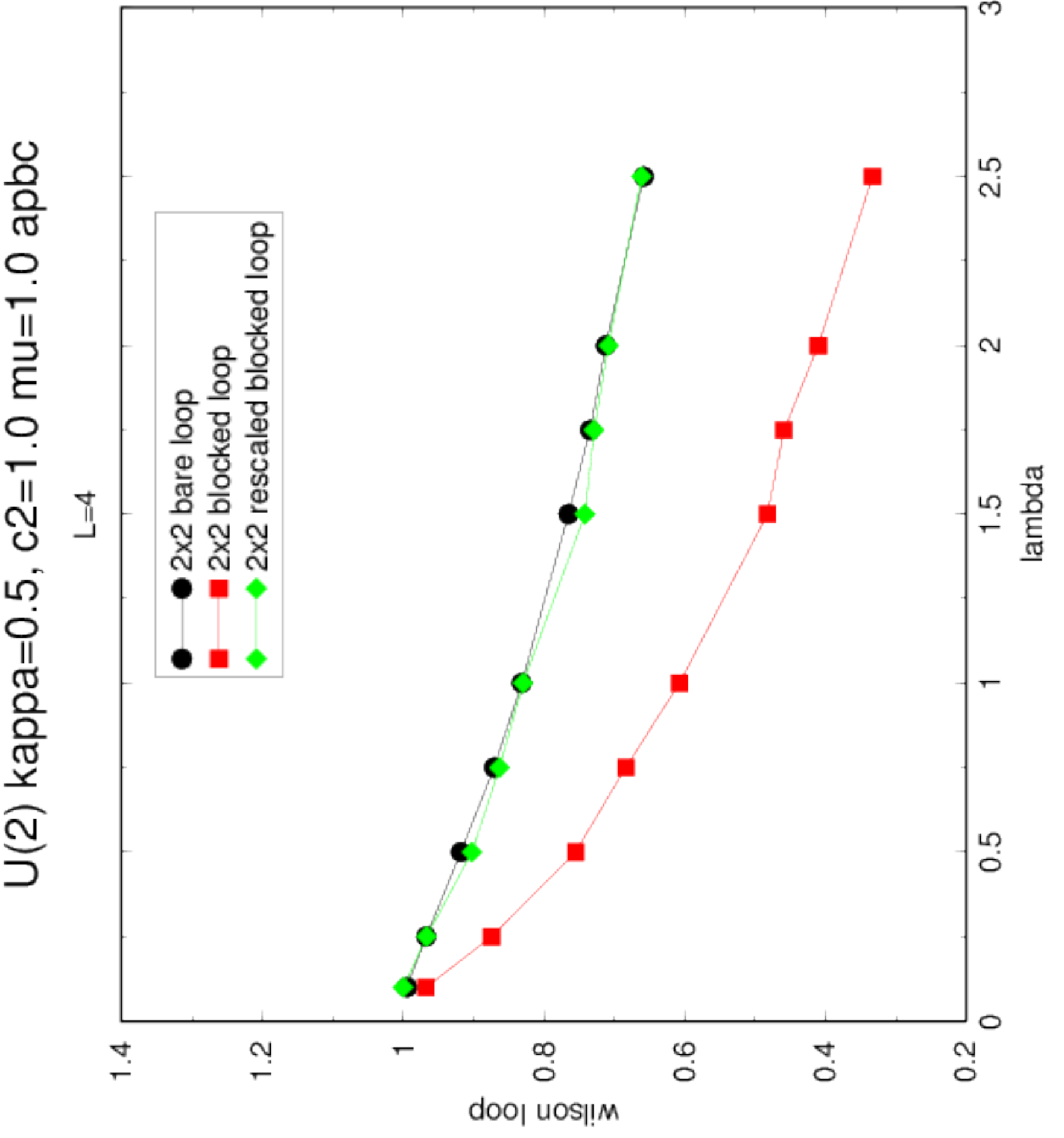}
\caption{A comparison of $W(2,2)_{b.f.}$ and $W(2,2)_{\text{coarse}}$ with
the rescaling factor taken into account.  \label{2b2} }
\end{center}
\end{figure}

As a preliminary step in this direction, we have performed a blocking
step $8^4 \to 4^4$.  The scaling parameter is determined from holding
the $1 \times 1$ Wilson loop constant.  I.e., if $W(1,1)_{b.f.} = \xi^4 W(2,2)_{\text{fine}}$
and $W(1,1)_{\text{coarse}}$ are set equal, then $\xi$ is determined.
Here ``b.f.'' indicates the fine lattice blocked using the RG blocking
transformations above.  We show the determination of this parameter in
Fig.~\ref{rescale}. These simulations utilize auxiliary parameters $\mu$ and $\kappa$ to regulate the flat directions and suppress the $U(1)$ sector [the gauge
group is $U(N)$ not $SU(N)$] --- we refer the reader to \cite{Catterall:2014vka} for details.  The current simulations employ $\mu=1.0$ and $\kappa=0.5$.
In addition, on both fine and coarse lattices the coupling $c2$ is set to its classical value $c2=1.0$ and the gauge coupling on both coarse and fine
lattices are set equal.  According to the above discussion, 
\beq
\xi^4 = \frac{W(1,1)_{\text{fine}}}{W(2,2)_{\text{coarse}}}
\eeq
Taking this rescaling into account for other Wilson loops, we show the
matching of $W(2,1)_{b.f.}$ and $W(2,2)_{b.f.}$ to $W(2,1)_{\text{coarse}}$
and $W(2,2)_{\text{coarse}}$ in Figs.~\ref{2b1} and \ref{2b2} respectively.
It can be seen that this simple rescaling factor is quite sufficient to
give a matching of Wilson loop observables.  There is no need to tune the coupling $c2$ on the coarse lattices to achieve
a good matching which implies that the system already lies close to a fixed point of the RG transformation. Of course in the
continuum this is to be expected, since the beta function vanishes for all gauge couplings, but it is quite
a startling result for the lattice theory we are studying.  One important point to make
about this result is that it suggests that the $\beta$ term
is not generated nonperturbatively, since we did not need to
add it to the coarse lattice theory in order to obtain matching.
Of course the current lattices are small, our statistics
are limited and the number and type  of operators used in the analysis is very small. We postpone a more detailed
analysis to a followup paper and regard the results presented here as merely a proof of principle for this new blocking scheme.

\section{Conclusions}
In this article we have exhibited a RG blocking scheme for $\Ncal=4$ lattice SYM that
preserves the symmetries and structure of the original lattice formulation:  $\cQ$ supersymmetry,
$S^5$ point group symmetry, $\eta$ shift symmetry, $U(N)$ gauge symmetry, the
hidden $U(1)$ ghost number, and the
spacetime realization of the fields in terms of 0-forms, 1-forms and 2-forms with
corresponding site, link, and diagonal gauge transformation properties.  The existence
of such a real space RG transformation is necessary to our arguments in \cite{Catterall:2011pd} about
the form of the long distance effective action, and the number of fine-tunings that are
required in order to recover the full symmetry group of the target continuum theory.

We have also shown that rescalings of the lattice fields reduces the number of
counterterms that must be adjusted in this procedure.  In addition, we have
argued that the so-called $\beta$ term lifts the moduli space, whereas the results of
\cite{Catterall:2011pd} prove that the moduli space is not lifted to all orders
in perturbation theory.  We therefore conclude that $\beta \equiv 0$, so that
there is one less fine-tuning.  Thus we finally arrive at a rather encouraging
result:  only a single parameter must be manipulated in order to obtain the desired
continuum limit.  This is  comparable to the tuning required in Wilson quark simulations of lattice QCD.

These results have led us to a preliminary implementation of MCRG.  We find that
using the rescaling freedom in the blocked link fields we are able to obtain
a matching of Wilson loops without any fine-tuning or flow of couplings at all.
This is consistent with an approximately conformal theory.

Follow-up work will include MCRG on larger lattices, and the inclusion of
matching observables that involve fermions.  This is important because
symmetry restoration must be checked in all sectors, not just the bosonic.
As mentioned above, our tests of $R_a$ symmetry in \cite{Catterall:2014vka} have been limited to
Wilson loops, and this is not a sufficient test to establish the full restoration of
$R_a$ symmetry, since fermionic observables should also be symmetric if the
lattice action is properly tuned.  It is somewhat surprising that \cite{Catterall:2014vka}
found an $\ord{10}$\% violation of the $R_a$ symmetry but that in the present
study we see no evidence for flow of couplings.  One possibility is that
the violation of $R_a$ symmetry is not having a significant effect on
conformality.  Another possibility is that $R_a$ symmetry tests are more
sensitive to deviations from the desired $\Ncal=4$ behavior.  

\section*{Acknowledgements}
Both SMC and JG received support from the
Department of Energy, Office of Science, Office of High Energy Physics,
under Grants Nos. SC0009998~and~DE-FG02-08ER41575 respectively. The numerical work utilized USQCD resources
at Fermilab. We gratefully acknowledge David Schaich for useful conversations and a careful reading of
the manuscript.

\appendix

\section{Wilson fermion action}
\label{wilson}
Here we enumerate the fine-tunings that would have to be performed
if Wilson fermions were used for the fermion discretization of lattice
${\cal N}=4$ SYM.
In the case of Wilson fermions, chiral symmetry is explicitly
broken by the regulator.  Thus one cannot preserve the $SU(4)_R$
of the continuum theory.  However, the $SO(4)$ subgroup can be
preserved.  Under this subgroup, the fermions $\lambda_i$, $i=1,\ldots,4$
(we use a two-component notation in terms of Weyl fermions)
transform as a ${\bf 4}$ and the scalars $\phi_m$, $m=1,\ldots,6$ transform
as a ${\bf 6}$, or antisymmetric representation, which we
can make explicity by mapping to $\phi_{ij}=-\phi_{ji}$, $i,j=1,\ldots,4$.
Then the most general long distance effective action consistent
with the symmetries of the lattice theory is
\beq
S &=& \int d^4 x ~ \tr \bigg\{ \frac{1}{2g_r^2} F_{\mu\nu} F_{\mu\nu} 
+ \frac{i}{g_r^2} \overline{\lambda}_i \overline{\sigma}^\mu D_\mu \lambda_i 
+ \frac{1}{g_r^2} D_\mu \phi_m D_\mu \phi_m + m_\phi^2 \phi_m \phi_m
\ddd  + m_\lambda ( \lambda_i \lambda_i + \overline{\lambda}_i \overline{\lambda}_i ) 
+ \kappa_1 \phi_m \phi_m \phi_n \phi_n + \kappa_2 \phi_m \phi_n \phi_m \phi_n 
+ y_1 ( \lambda_i [ \phi_{ij} , \lambda_j ] 
+ \overline{\lambda}_i [\phi_{ij}, \overline{\lambda}_j ] )
\ddd + y_2 \epsilon_{ijkl} ( \lambda_i [ \phi_{jk} , \lambda_l ] 
+ \overline{\lambda}_i [\phi_{jk}, \overline{\lambda}_l ] ) \bigg\}
\ddd + \int d^4 x ~ \bigg\{ \kappa_3 ( \tr \phi_m \phi_m)^2 
+ \kappa_4 \tr \phi_m \phi_n \tr \phi_m \phi_n \bigg\}
\eeq
The coefficients of the first three terms were achieved by rescaling the fields.
The other eight coefficients will be determined by the renormalization group
flow, and must be fine-tuned by adjusting corresponding UV coefficients in
the lattice theory.

\section{Continuum limit of the $\beta$ term}
\label{beta}
To arrive at the continuum limit of the $\beta$ term discussed in the
main text, we apply the link expansion \myref{linkexpansion} and
keep the terms that are not $\ord{a}$ suppressed.  This leads to
\beq
&& \int d^4 x ~ \beta \frac{1}{a} \bigg\{ \sum_a \tr ( d (\cA_a - \cAb_a) ) 
- \frac{1}{N} \tr d \sum_a \tr ( \cA_a - \cAb_a )
\ddd \qquad - \sum_a \tr \eta \psi_a + \frac{1}{N} \tr \eta \sum_a \tr \psi_a + \ord{a} \bigg\}
\label{vfex}
\eeq
Now we recall that
\beq
\cA_a = A_a + i B_a, \quad \cAb_a = A_a - i B_a
\label{Aex}
\eeq
where $A_a$ gives rise to the ordinary gauge fields and one scalar,
and $B_a$ lead to the other five scalars of $\Ncal=4$ SYM.
Thus $\cA_a - \cAb_a = 2i B_a$.  Also we decompose the $U(N)$ generators into
$T^0 = (i/\sqrt{2N}) {\bf 1}_N$ and $T^A \in su(N)$ with $A = 1,\ldots , N^2-1$.  We will
normalize the SU(N) generators to $\tr T^A T^B = (1/2) \delta^{AB}$.  What
we find is that all of the U(1) fields disappear from the above expression
and we are left with:
\beq
\int d^4 x ~ \frac{\beta}{2} \frac{1}{a} \bigg\{ 2i d^A \sum_a B_a^A
- \eta^A \sum_a \psi_a^A + \ord{a} \bigg\}
\label{sufex}
\eeq
Since the mass dimension of the auxiliary field is $[d]=2$, what we see is
that we have two dimension three operators, with a coefficient with mass
dimension one.  Absent fine-tuning (or our moduli space argument), the size
of this coefficient is $\ord{1/a}$.

It is now desirable to eliminate the auxiliary field.  For this we need
all of the terms in the action that involve $d$.  We evaluate
\beq
[ \cDb_a , \cD_a ] = 2 i D_a B_a = 2 i ( \p_a B_a + [ A_a , B_a ] )
\eeq
Then the terms in the action with the auxiliary field are
\beq
\tr \( 2 i \alpha_2 d D_a B_a - \frac{\alpha_3}{2} d^2 \)
+ \frac{\beta}{2a} 2 i d^A \sum_a B_a^A
\label{auxact}
\eeq
Solving the auxiliary equations of motion yields
\beq
d^0 &=&  2i \frac{\alpha_2}{\alpha_3} \p_a B_a^0 \nnn
d^A &=&  2i \frac{\alpha_2}{\alpha_3} (D_a B_a)^A + 2i \frac{\beta}{\alpha_3} \frac{1}{a} \sum_a B_a^A
\eeq
Substituting these back into \myref{auxact} yields
\beq
&& -\frac{\alpha_2^2}{\alpha_3} ( \p_a B_a^0 )^2 - \frac{\alpha_2^2}{\alpha_3} (D_a B_a)^A (D_b B_b)^A
- \frac{2 \alpha_2 \beta}{\alpha_3} \frac{1}{a} (D_a B_a)^A \sum_b B_b^A
\ddd \qquad - \frac{\beta^2}{\alpha_3} \frac{1}{a^2} \sum_a B_a^A \sum_b B_b^A
\eeq
Thus we see that the SU(N) scalar mode $\sum_a B_a^A$ gets an $\ord{1/a^2}$ mass term.
In addition, we have a cubic interaction $[\p_a + A_a, B_a ] \sum_b B_b$ in the
SU(N) sector.  Both of these would lift the moduli space, which we have shown
previously in \cite{Catterall:2011pd} does not occur to any order in perturbation 
theory.  Thus unless nonperturbative
effects lift the moduli space, $\beta \equiv 0$.  (It can also be seen from \myref{sufex} that
for $\beta\not=0$ the SU(N) fermions would get a mass term $\eta \sum_a \psi_a$.)

\clearpage

\bibliography{rsrg}
\bibliographystyle{JHEP}

\end{document}